\documentclass[12pt]{article}

\usepackage{graphicx}
\usepackage{amsmath}
\usepackage{amssymb}
\usepackage{bm}

\numberwithin{equation}{section}

\usepackage{color}
\input{colordvi.tex}

\allowdisplaybreaks

\newcommand{\vac}[1]{\left|{0}\right>}

\begin{document}

\begin{titlepage}
\thispagestyle{empty}
\begin{flushleft}
\hfill September, 2012
\end{flushleft}

\vskip 1.5 cm
\bigskip

\begin{center}
{\Large\bf{
On Nontrivial Solutions around a Marginal Solution
}}
\vskip 1em
{\Large\bf{
in Cubic Superstring Field Theory
}}\\

\renewcommand{\thefootnote}{\fnsymbol{footnote}}

\vskip 2.5cm
{\large
Shoko {\sc Inatomi}$^{1}$,
Isao {\sc Kishimoto}$^{2}$ 
and Tomohiko {\sc Takahashi}$^{1}$}\\

\bigskip\bigskip

{\it
$^{1)}$ Department of Physics, Nara Women's University,
Nara 630-8506, Japan,\\
$^{2)}$ Faculty of Education, Niigata University,
Niigata 950-2181, Japan
}
\vskip 2in
\end{center}
\begin{abstract}
We construct tachyon vacuum and half-brane solutions,
using an extension of $KBc$ algebra, in the theory around a type of
identity-based marginal solutions in modified cubic superstring field
 theory.
With explicit computations,
we find that their vacuum energies are the same as those of
 corresponding solutions around the original theory.
It implies that the vacuum energy for the identity-based marginal
 solution vanishes although straightforward computation of it  is
 subtle.
We also evaluate the gauge invariant overlaps for those nontrivial
 solutions.
The values for them are deformed according to the marginal solution in
 the same way as the case of bosonic string field theory.

\end{abstract}
\end{titlepage}\vfill\setcounter{footnote}{0} \renewcommand{\thefootnote}{
\arabic{footnote}} \newpage

\section{Introduction
\label{sec:intro}}

In a recent development in string field theory (SFT), so called $KBc$
algebra \cite{Okawa:2006vm} is often used because of its algebraic
simplicity in order to construct classical solutions and evaluate gauge
invariants.
In \cite{Inatomi:2012nv}, the authors have investigated ``$K'Bc$ algebra,''
where the Kato-Ogawa BRST operator $Q_{\rm B}$ and a string field
$K=\{Q_{\rm B},B\}$ are
replaced by $Q'$ and $K'=\{Q',B\}$, respectively, in $KBc$ algebra.
Here, $Q'$ is the BRST operator in a theory around a class of identity-based
marginal solutions \cite{Takahashi:2001pp,
Takahashi:2002ez, Kishimoto:2005bs} in bosonic open SFT.
Using $K'Bc$ algebra, the tachyon vacuum solution on the deformed background
was constructed as in \cite{Erler:2009uj} and vacuum energy and gauge
invariant overlap are evaluated explicitly. It turned out that the value
of vacuum energy does not change.
But the expression of the gauge invariant overlap is deformed
appropriately. In particular, for a closed tachyon vertex,
a phase factor appears and it is the same value as that evaluated in
\cite{Katsumata:2004cc} using a different method.

In this paper, we extend computations performed in \cite{Inatomi:2012nv} to the NS
sector of modified cubic superstring field theory (SSFT)
\cite{Preitschopf:1989fc, Arefeva:1989cp, Arefeva:1989cm} without GSO
projection.
In the case of superstring, $KBc$ algebra with a string field $\gamma$
was applied to a construction of tachyon vacuum
solution \cite{Erler:2007xt} and its variants \cite{Aref'eva:2000mb,
Aref'eva:2009ac, Gorbachev:2010zz, Arroyo:2010fq, AldoArroyo:2012if}.
It was further extended to ``$GKBc\gamma$ algebra'' by including 
a string field $G$, such as $G^2=K$, made of superconformal
generators and then
 half-brane solutions, whose vacuum energy is half of that of
the tachyon vacuum solution,  were constructed \cite{Erler:2010pr}.
In this context, we explore $G'K'Bc\gamma$ algebra,
where $Q_{\rm B}$, $K$ and $G$ are replaced by $Q'$, $K'=\{Q',B\}$ and
$G'$ (such as $G'^2=K'$), respectively,
corresponding to $K'Bc$ algebra in the bosonic case \cite{Inatomi:2012nv}.
Here, $Q'$ is the BRST operator in a theory around a class of
identity-based marginal solutions in modified cubic SSFT
\cite{Kishimoto:2005bs}.\footnote{
In SSFT, other types of marginal solutions based on wedge states are also
constructed \cite{Erler:2007rh,Okawa:2007ri,Okawa:2007it,Fuchs:2007gw,Kiermaier:2007ki},
but identity-based solutions are suitable for our purpose.
}
Around an identity-based marginal solution $\Psi_J$,
which is constructed using a supercurrent algebra, we have an explicit
expression of a deformed BRST operator $Q'$.
Using $G'K'Bc\gamma$ algebra, we can construct various solutions
to the equation of motion, $\hat Q'\Phi+\Phi^2=0$,  in the NS sector
without GSO projection.
In particular, we focus on a tachyon vacuum solution and a half-brane
solution in the marginally deformed background and we evaluate the
vacuum energy and gauge invariant overlaps for them.
The results are similar to the bosonic case:
The vacuum energies are the same as those for corresponding 
solution in the original theory.
The values of the gauge invariant overlap with a closed tachyon vertex
change by a phase factor just as in \cite{Inatomi:2012nv}.
We find that the vacuum energy and the gauge invariant overlap for the
half-brane solution in the marginally deformed background
are half of those for the tachyon vacuum solution, respectively.
These results are consistent with our expectation that the vacuum energy
for the identity-based solution $\Psi_J$ vanishes and it corresponds to a
marginal deformation.

This paper is organized as follows. 
In the next section, we will develop $G'K'Bc\gamma$ algebra in a
theory around a particular identity-based marginal solution.
In \S\ref{sec:energy} and \S\ref{sec:overlap}, we will calculate vacuum
energy and gauge invariant overlap for a tachyon vacuum 
solution and a half-brane solution.
Then, we will give some concluding remarks in \S\ref{sec:remarks}.
In appendix \ref{sec:convention}, we will present our conventions
including a brief review of $GKBc\gamma$ algebra. In appendices
\ref{sec:expansion} and \ref{sec:Jindep}, we will show technical
details. 
In appendix \ref{sec:fieldredefinition}, we will discuss
the gauge invariant overlap with a closed tachyon vertex and a field
redefinition induced by the identity-based marginal solution as in
\cite{Katsumata:2004cc}.

\section{A version of $GKBc\gamma$ algebra and classical solutions in a
 marginally deformed background
\label{sec:KBc}}

Here, we give a version of ``$GKBc\gamma$ algebra''
developed in \cite{Erler:2010pr}\footnote{
See appendix \ref{sec:convention} for a review and our conventions.}
in the theory around an identity-based marginal solution. 
Using this algebra, we can construct classical solutions including
GSO$(-)$ sector easily as in the original theory.

First of all, we consider a theory around a particular identity-based
solution $\Psi_J$, which is given by
\begin{eqnarray}
&&\Psi_J=-V^a_L(F_a)I
+\frac{1}{8}\Omega^{ab}C_L(F_aF_b)I,
\label{eq:PsiJdef}
\\
&&V^a_L(f)\equiv\int_{C_{\rm L}}\frac{dz}{2\pi
 i}\frac{1}{\sqrt{2}}f(z)(cJ^a(z)+\gamma \psi^a(z)),
~~~C_L(f)\equiv\int_{C_{\rm L}}\frac{dz}{2\pi i}f(z)c(z),
\end{eqnarray}
where $F_a(z)$ is some function such as $F_a(-1/z)=z^2F_a(z)$,
$C_{\rm L}$ denotes a half unit circle: $|z|=1,{\rm Re}\,z\ge 0$
and 
$I$ is the identity state. 
In (\ref{eq:PsiJdef}), the repeated indices, $a$, $b$, are contracted.
The above $\Psi_J$ satisfies the equation of
motion to modified cubic SSFT in the NS sector because we can show
\begin{eqnarray}
 Q_{\rm B}\Psi_J+\Psi_J*\Psi_J=0.
\end{eqnarray}
Actually, $\Psi_J=e^{-\Phi_J}Q_{\rm B}e^{\Phi_J}$ holds, where
$\Phi_J=-\tilde{V}^a_L(F_a)I$ is an identity-based marginal solution
to Berkovits' WZW-type SSFT investigated 
in \cite{Kishimoto:2005bs}.
Here, we suppose that ${\bm J}^a(z,\theta)=\psi^a(z)+\theta J^a(z)$ is a
supercurrent associated with a Lie algebra, where $a$ is its index, in
the matter sector.
Its component fields satisfy following 
operator product expansions (OPE)
\cite{Mohammedi:1993rg}:
\begin{eqnarray}
&&\psi^a(y)\psi^b(z)\sim \frac{1}{y-z}\frac{1}{2}\Omega^{ab},
~~~~J^a(y)\psi^b(z)\sim \frac{1}{y-z}f^{ab}_{~~\,c}\psi^c(z),\\
&&J^a(y)J^b(z)\sim
  \frac{1}{(y-z)^2}\frac{1}{2}\Omega^{ab}+\frac{1}{y-z}f^{ab}_{~~\,c}J^c(z),
\label{eq:JJ_OPE}
\end{eqnarray}
where constants $\Omega^{ab},f^{ab}_{~~\,c}$ satisfy following relations
\begin{eqnarray}
&&\Omega^{ab}=\Omega^{ba},
~~~~~~f^{ab}_{~~\,c}\Omega^{cd}+f^{ad}_{~~\,c}\Omega^{cb}=0,
\label{eq:f_Rel1}
\\
&&f^{ab}_{~~\,c}=-f^{ba}_{~~\,c},
~~~~f^{ab}_{~~\,d}f^{cd}_{~~\,e}
+f^{bc}_{~~\,d}f^{ad}_{~~\,e}+f^{ca}_{~~\,d}f^{bd}_{~~\,e}=0.
\label{eq:f_Rel2}
\end{eqnarray}

By re-expanding the NS action of modified cubic SSFT around the solution
$\Psi_J$ (\ref{eq:PsiJdef}), the BRST operator is deformed as
\begin{eqnarray}
 Q'&=&Q_{\rm B}+ [\Psi_J,\,\cdot\,\}_*\nonumber\\
&=&Q_{\rm B}-V^a(F_a)+\frac{1}{8}\Omega^{ab}C(F_aF_b),
\label{eq:Q'def}
\end{eqnarray}
where $V^a(F_a)$ and $C(F_aF_b)$ are given by integrations
along the whole unit circle, $|z|=1$:
\begin{eqnarray}
&&V^a(f)\equiv\oint\frac{dz}{2\pi
 i}\frac{1}{\sqrt{2}}f(z)(cJ^a(z)+\gamma \psi^a(z)),
~~~C(f)\equiv\oint\frac{dz}{2\pi i}f(z)c(z).
\end{eqnarray}
The superconformal generators, $L'_n,G'_r$, corresponding to the above $Q'$
(\ref{eq:Q'def}) are given by
\begin{eqnarray}
 L'_n&\equiv&\{Q',b_n\}=L_n
-\frac{1}{\sqrt{2}}\sum_{k\in {\mathbb{Z}}}F_{a,k}J^a_{n-k}
+\frac{1}{8}\Omega^{ab}\sum_{k\in{\mathbb{Z}}}F_{a,n-k}F_{b,k},\\
G'_r&\equiv &[Q',\beta_r]=G_r-\frac{1}{\sqrt{2}}
\sum_{k\in {\mathbb{Z}}}F_{a,k}\psi^a_{r-k},
\end{eqnarray}
where we define the coefficients  as $F_{a,n}\equiv \oint
 \frac{d\sigma}{2\pi}e^{i(n+1)\sigma}F_a(e^{i\sigma})$
and then $F_a(-1/z)=z^2F_a(z)$ implies $F_{a,n}=-(-1)^nF_{a,-n}$.
Only the matter sectors of them are deformed and the central charge is
not changed.
Replacing  $L_n$, $G_r$ in $K_1^L$ (\ref{eq:K1Ldef}) and ${\cal G}_L$
(\ref{eq:calG_L}) with $L'_n$, $G'_r$, respectively, we define
 $K_1^{\prime L}$, ${\cal G}'_L$ and  string fields $K',G'$:
\begin{eqnarray}
 K'=\frac{\pi}{2}K_1^{\prime L}I,~~~~G'={\cal G}'_LI\sigma_1
\end{eqnarray}
as in (\ref{eq:KBcdef}), (\ref{eq:Ggammadef}). Here, $\sigma_1$ 
in $G'$ is a Chan-Paton factor, which is involved to treat a theory without GSO
projection.
Because of  $Q'|I\rangle=0$, we have
\begin{eqnarray}
&&K'_1|I\rangle=(Q'B_1+B_1Q')|I\rangle=0,\\
&&{\cal G}'|I\rangle=\sqrt{\frac{\pi}{2}}
(Q'\tilde\beta_{-\frac{1}{2}}-\tilde\beta_{-\frac{1}{2}}Q')|I\rangle=0,
\end{eqnarray}
where $K_1'$ and ${\cal G}'$ are defined by replacing 
$L_n,G_r$ with $L'_n,G'_r$ in $K_1=\{Q_{\rm B},B_1\},{\cal
G}=\sqrt{\frac{\pi}{2}}[Q_{\rm B},\tilde\beta_{-\frac{1}{2}}]$,
respectively.
The above relations are consistent with a deformed version of 
(\ref{eq:partialdel}):
\begin{eqnarray}
 \partial' \Phi\equiv K'\Phi-\Phi K'=\frac{\pi}{2}K_1'\Phi,~~~~
\delta' \Phi\equiv G'\Phi-(-)^{F(\Phi)}\Phi G'=({\cal G}'\sigma_1)\Phi,
\end{eqnarray}
where $(-)^{F(\Phi)}$ is a sign factor due to worldsheet spinor.
Attaching a Chan-Paton factor $\sigma_3$ and using a notation $\hat
Q'\equiv Q'\sigma_3$, we have a deformed version
of $GKBc\gamma$ algebra reviewed in appendix \ref{sec:convention}:
\begin{eqnarray}
&&G'^2=K',~~~\hat Q'B=K',~~~\hat Q'K'=0,~~~\hat Q' G'=0,\\
&&\hat Q'c=cK'c-\gamma^2=cKc-\gamma^2,\\
&&BG'=G'B,~~~~BK'=K'B,~~~~G'K'=K'G',\\
&&\delta' c=2i\gamma,~~~~\delta' \gamma=-\frac{i}{2}\partial' c=
-\frac{i}{2}\partial c,~~~~
\delta' G'=2K',~~~\delta' K'=0,~~~\delta' B=0.~~~~
\end{eqnarray}
We also note that
\begin{eqnarray}
 \hat Q'\gamma=\hat Q\gamma=c\partial \gamma-\frac{1}{2}(\partial
  c)\gamma,
~~~~c\partial \gamma=-(\partial\gamma) c,~\gamma\partial c=-(\partial
c)\gamma.
\end{eqnarray}
In the following, we call the above relations among string fields as 
$G'K'Bc\gamma$ algebra, which is the same form as $GKBc\gamma$
algebra. We will use the above relations extensively in various
calculations.

Let us consider the action $S'[\Phi]$, which is obtained by
re-expanding around $\Psi_J\sigma_3$, where $\sigma_3$ is the
Chan-Paton factor for GSO$(+)$ string field (\ref{eq:Psi_++Psi_-})
 in the NS action without GSO projection (\ref{eq:action}).
More explicitly, it is defined by
\begin{eqnarray}
S'[\Phi]&=&S[\Phi+\Psi_J\sigma_3]-S[\Psi_J\sigma_3]\nonumber\\
&=&\frac{1}{2}\langle\!\langle
\Phi\,\hat Q'\Phi\rangle\!\rangle
+\frac{1}{3}\langle\!\langle\Phi^3\rangle\!\rangle.
\label{eq:action'}
\end{eqnarray}
The equation of motion of $S'[\Phi]$ is\footnote{
In this paper, we ignore the kernel of the picture changing operator
with picture number $(-2)$, $Y_{-2}$, for simplicity.
See \cite{Kohriki:2012pp} for a recent argument.
}
\begin{eqnarray}
 \hat Q'\Phi+\Phi^2=0.
\label{eq:EOM'}
\end{eqnarray}
Using the method in \cite{Erler:2010pr}
with $G'K'Bc\gamma$ algebra instead of $GKBc\gamma$ algebra, we can easily
construct a class of solutions to (\ref{eq:EOM'}):
\begin{eqnarray}
&&\Phi_{f'}=\sqrt{f'}\left(c\frac{K'B}{1-f'}c+B\gamma^2\right)\sqrt{f'}
=\sqrt{f'}\left(c\frac{K'f'}{1-f'}Bc+\hat Q'(Bc)\right)\sqrt{f'},~~~
\label{eq:Phif'}
\end{eqnarray}
where $f'$ is a function of $G'$, noting  $K'=G'^2$.

In the following sections, we consider two solutions, 
which correspond to 
$f'=\frac{1}{1+K'}$ (a tachyon vacuum solution $\Phi_T$) and
$f'=\frac{1}{1+iG'}$ (a half-brane solution $\Phi_H$).
More explicitly, they are given by
\begin{eqnarray}
 \Phi_T&=&\frac{1}{\sqrt{1+K'}}\left(c+\hat Q'(Bc)
\right)
\frac{1}{\sqrt{1+K'}},
\label{eq:Phi_T}
\\
 \Phi_H&=&\frac{1}{\sqrt{1+iG'}}\left(
-icG'Bc+\hat Q'(Bc)
\right)
\frac{1}{\sqrt{1+iG'}}.
\label{eq:Phi_H}
\end{eqnarray}

\section{Vacuum energy of the solutions on the marginally deformed
 background\label{sec:energy}}

In this section, we evaluate the vacuum energy, or the value of the
action, for solutions $\Phi_T$ (\ref{eq:Phi_T}) and $\Phi_H$
(\ref{eq:Phi_H}) in the theory
given by the marginally deformed action (\ref{eq:action'}).
In general, for a solution $\Phi_{f'}$
(\ref{eq:Phif'}) to the equation of motion
(\ref{eq:EOM'}), the value of the action is computed as
\begin{eqnarray}
&&S'[\Phi_{f'}]=\frac{1}{6}\langle\!\langle
\Phi_{f'}\,\hat Q'\Phi_{f'}\rangle\!\rangle
=\frac{1}{6}\langle\!\langle
c\frac{K'f'}{1-f'}Bcf'\hat{Q}'\!\left(c
\frac{K'f'}{1-f'}Bc\right)\!f'
\rangle\!\rangle,
\label{eq:S'Phif'}
\end{eqnarray}
using $G'K'Bc\gamma$ algebra and a deformed version of
(\ref{eq:hatQPhiPsi}) and (\ref{eq:hatQ()}):
\begin{eqnarray}
&&\hat Q'(\Phi\Psi)=(\hat
 Q'\Phi)\Psi+(-1)^{\epsilon(\Phi)+F(\Phi)}\Phi(\hat Q'\Psi),\\
&&\langle\!\langle\hat Q'(\cdots )\rangle\!\rangle=0,
\end{eqnarray}
where $(-)^{\epsilon(\Phi)}$ is a sign factor from Grassmannality
of $\Phi$.

\subsection{Tachyon vacuum solution}

In the case of $f'=\frac{1}{1+K'}$, namely, $\Phi_T$ (\ref{eq:Phi_T}),
the expression of (\ref{eq:S'Phif'}) is simplified as
\begin{eqnarray}
 S'[\Phi_T]=\frac{1}{6}\langle\!\langle
c\frac{1}{1+K'}(c\partial c-\gamma^2)
\frac{1}{1+K'}
\rangle\!\rangle
=-\frac{1}{6}\langle\!\langle
\gamma^2\frac{1}{1+K'}c
\frac{1}{1+K'}
\rangle\!\rangle.
\label{eq:S'Phi_T1}
\end{eqnarray}
In the second equality, we have used the form of the picture changing
operator $Y_{-2}$ defined in (\ref{eq:Y_-2def}).

In order to define the inverse of a string field $I+K'$,
which is denoted by $\frac{1}{1+K'}$, we use the following expression:
\begin{eqnarray}
&&\frac{1}{1+K'}=\int_0^{\infty}dt\, e^{-t(1+K')}.
\end{eqnarray}
Here, $K'$ can be rewritten as
\begin{eqnarray}
&&K'=K-J+\frac{\pi}{2}{\cal C}I,\\
&&J=\frac{\pi}{2}\int_{-\infty}^{\infty}dt f_a(t)\hat{U}_1\tilde
 J^a(it)|0\rangle,~~~~~f_a(t)\equiv
 \frac{F_a(\tan(it+\frac{\pi}{4}))}{2\pi\sqrt{2}\cos^2(it+\frac{\pi}{4})},
\\
&&{\cal C}=
\int_{C_{\rm L}}\frac{dz}{2\pi i}(1+z^2)\frac{\Omega^{ab}}{8}F_a(z)F_b(z)
=\frac{\pi}{2}\int_{-\infty}^{\infty}dt
\Omega^{ab}f_a(t)f_b(t).
\label{eq:calCdef}
\end{eqnarray}
($\tilde J^a(\tilde z)=(\cos\tilde z)^{-2}J^a(\tan\tilde z)$ 
is $J^a$ in the sliver frame.)
Using the expansion formula (\ref{eq:eXdelXN}) and the methods developed in
\cite{Schnabl:2005gv},
the order $J^N$ term of $e^{-tK+tJ}$ is computed as
\begin{eqnarray}
&&e^{-tK+tJ}|_{O(J^N)}
\nonumber\\
&&=
\int_0^1\!du_1\!\int_0^{1-u_1}\!\!\!\!\!du_2\cdots \!
\int_0^{1-u_1-u_2\cdots -u_{N-1}}\!\!\!\!\!\!\!\!du_N t^N\!
e^{-t(1-u_1-u_2\cdots -u_N)K} J e^{-tu_1K} J
e^{-t u_2 K}\! \cdots J e^{-tu_NK}
\nonumber\\
&&=t^N\!\!
\int_0^1\!du_1\!\int_0^{1-u_1}\!\!\!\!\!du_2\cdots \!\!
\int_0^{1-u_1-u_2\cdots -u_{N-1}}\!\!\!\!\!\!\!\!\!du_N\!
\int_{-\infty}^{\infty}\!\!dt_1f_{a_1}(t_1)\!
\int_{-\infty}^{\infty}\!\!dt_2f_{a_2}(t_2)\cdots\! 
\int_{-\infty}^{\infty}\!\!dt_Nf_{a_N}(t_N)
\nonumber\\
&&~~~~~~~\times 
\frac{\pi^N}{2^N}\hat U_{t+1}\tilde
J^{a_1}(it_1+\frac{\pi}{4}y_1)
\tilde
J^{a_2}(it_2+\frac{\pi}{4}y_2)
\cdots
\tilde
J^{a_N}(it_N+\frac{\pi}{4}y_N)|0\rangle,
\label{eq:exp1}
\end{eqnarray}
where in the real part of the argument of $\tilde J^{a_i}$, $y_i$
($i=1,2,\cdots, N$) are defined as
\begin{eqnarray}
&&y_1=2t\sum_{k=1}^Nu_k-t,~\cdots,~~
y_i=2t\sum_{k=i}^Nu_k-t,~\cdots,~~
y_N=2tu_N-t.
\end{eqnarray}
Hence, $u_i$ can be expressed using $y_i$ and the Jacobian is computed as
\begin{eqnarray}
&&u_1=\frac{1}{2t}(y_1-y_2),~\cdots,~~u_i=\frac{1}{2t}(y_i-y_{i+1}),~\cdots,
~~u_N=\frac{1}{2t}y_N+\frac{1}{2},\\
&&
 \frac{\partial(u_1,\cdots,u_N)}{\partial(y_1,\cdots,y_N)}=2^{-N}t^{-N}.
\end{eqnarray}
Using the above, the integration in (\ref{eq:exp1}) can be rewritten as
\begin{eqnarray}
&&e^{-tK+tJ}|_{O(J^N)}
\nonumber\\
&&=\frac{\pi^N}{4^N}\int_{-t}^t\!dy_1\!\int_{-t}^{y_1}\!\!dy_2\cdots \!
\int_{-t}^{y_{N-1}}\!\!\!dy_N
\int_{-\infty}^{\infty}\!\!dt_1f_{a_1}(t_1)
\int_{-\infty}^{\infty}\!\!dt_2f_{a_2}(t_2)\cdots 
\int_{-\infty}^{\infty}\!\!dt_Nf_{a_N}(t_N)
\nonumber\\
&&~~~~~\times 
\hat U_{t+1}\tilde
J^{a_1}(it_1+\frac{\pi}{4}y_1)
\tilde
J^{a_2}(it_2+\frac{\pi}{4}y_2)
\cdots
\tilde
J^{a_N}(it_N+\frac{\pi}{4}y_N)|0\rangle.
\end{eqnarray}
If we use an ordering symbol ${\bf T}$
with respect to the real part of the arguments of
$\tilde J^{a_i}$ in the above integrations, we have
\begin{eqnarray}
&&e^{-tK+tJ}|_{O(J^N)}
=\frac{1}{N!}\hat U_{t+1}\,
{\bf T}
\left(
\frac{\pi}{4}\int_{-t}^t du \int_{-\infty}^{\infty}dt'
f_a(t')
\tilde J^a(it'+\frac{\pi}{4}u)\right)^N|0\rangle,
\end{eqnarray}
which implies the following expression of
$e^{-tK'}=e^{-t(K-J+\frac{\pi}{2}{\cal C}I)}$:
\begin{eqnarray}
 e^{-tK'}&=&e^{-t\frac{\pi}{2}{\cal C}}
\hat U_{t+1}{\bf T}\exp\left(
\frac{\pi}{4}\int_{-t}^t du \int_{-\infty}^{\infty}dt'
f_a(t')
\tilde J^a(it'+\frac{\pi}{4}u)\right)|0\rangle.
\label{eq:e-tKp}
\end{eqnarray}
Therefore, (\ref{eq:S'Phi_T1}) is computed as
\begin{eqnarray}
 S'[\Phi_T]
=-\frac{1}{6}\int_0^{\infty}dt\int_0^{\infty}ds\, e^{-t-s}
\langle\!\langle
\gamma^2 e^{-tK'} c e^{-sK'}
\rangle\!\rangle,
\label{eq:S'Phi_T2}
\end{eqnarray}
where the integrand can be rewritten as
\begin{eqnarray}
&&\langle\!\langle \gamma^2 e^{-tK'}ce^{-sK'}\rangle\!\rangle
\nonumber\\
&&=\frac{4}{\pi^2}
\langle I|Y_{-2}\hat {U}_{t+s+1}\tilde\gamma^2(\frac{\pi}{4}(t+s))
\tilde c(\frac{\pi}{4}(s-t))
\nonumber\\
&&~~~~~~~~~\times
e^{-(t+s)\frac{\pi}{2}{\cal C}}\,{\bf T}\exp\left(
\frac{\pi}{4}\int_{-t-s}^{t+s}\!\!du
\int_{-\infty}^{\infty}\!\!\!dt' f_a(t')
\tilde J^a(it'\!+\!\frac{\pi}{4}u)\right)\!|0\rangle\nonumber\\
&&=\frac{(t+s)^2}{\pi^2}
\langle 0|\tilde Y(i\infty)\tilde Y(-i\infty)
\tilde\gamma^2(\frac{\pi}{2})
\tilde c(\frac{\pi(s-t)}{2(t+s)})
\nonumber\\
&&~~~~~~~~~~~~~~~\times
e^{-(t+s)\frac{\pi}{2}{\cal C}}\,{\bf T}\exp\left(
\int_{-\frac{\pi}{2}}^{\frac{\pi}{2}}\!\!du
\int_{-\infty}^{\infty}\!\!\!dt' f_a(t')
\tilde J^a(\frac{2it'}{t+s}+u)\right)\!|0\rangle,
\end{eqnarray}
using (\ref{eq:e-tKp}), $\hat U_1\hat U_{r+1}=\hat U_r$
and $\lambda^{{\cal L}_0}\tilde \phi (\tilde z)
\lambda^{-{\cal L}_0}
=\lambda^h \tilde \phi(\lambda\tilde z)
$ for a primary field with dimension $h$ in the sliver frame.
Furthermore, with (\ref{eq:Y_-2def}) 
and\footnote{
We note that
\begin{eqnarray}
 \langle
  \delta'(\gamma(w))\delta'(\gamma(z))\gamma(y_1)\gamma(y_2)\rangle
=-\frac{1}{(w-z)^3}\left(2wz+2y_1y_2-(w+z)(y_1+y_2)\right),
\end{eqnarray}
in the $\beta\gamma$-sector, which can be derived from the
$\xi\eta\phi$-expression.
}
\begin{eqnarray}
\langle \tilde c(iM)\tilde c(-iM)
\tilde c(z)\rangle
&\sim&\frac{i}{8}e^{4M},
~~~~(M\to +\infty)\\
\langle \delta'(\tilde\gamma(iM))\delta'(\tilde\gamma(-iM))\tilde \gamma(x)
\tilde\gamma(y)\rangle
&\sim& -i\frac{4}{e^{4 M}}\cos(x-y),~~~~(M\to +\infty)
\end{eqnarray}
in the ghost sector, we have
\begin{eqnarray}
&&\langle\!\langle \gamma^2 e^{-tK'}ce^{-sK'}\rangle\!\rangle
\nonumber\\
&&=-\frac{(t+s)^2}{\pi^2}\lim_{M\to\infty}\langle 
\delta'(\tilde \gamma(iM))\delta'(\tilde \gamma(-iM))
\tilde \gamma^2(\frac{\pi}{2})\rangle_{\beta\gamma}
\langle \tilde c(iM)\tilde c(-iM)\tilde c(\frac{\pi(s-t)}{2(s+t)})
\rangle_{bc}
\nonumber\\
&&~~~~~~~~~~~~~~~\times e^{-(t+s)\frac{\pi}{2}{\cal C}}
\left\langle \exp\left(
\int_{-\frac{\pi}{2}}^{\frac{\pi}{2}}\!\!du
\int_{-\infty}^{\infty}\!\!\!dt' f_a(t')
\tilde J^a(\frac{2it'}{t+s}\!+\!u)\right)\right\rangle_{\rm mat}
\nonumber\\
&&
=-\frac{(t+s)^2}{\pi^2}\lim_{M\to\infty}(-4ie^{-4M})\frac{i}{8}e^{4M}\cdot
1
=-\frac{(t+s)^2}{2\pi^2}.
\label{eq:evalTacpre}
\end{eqnarray}
The minus sign comes from Grassmannality of $\delta'(\gamma)$,
which corresponds to $\partial\xi e^{-2\phi}$
in the $\xi\eta\phi$-expression.
In the first equality, a ${\bf T}$-ordered exponential 
becomes a conventional exponential in the CFT correlator and 
in the second equality, we have used (\ref{eq:mat=1}) in the matter
sector, which is essential for $J$-independence.

Using this result, the value of the action (\ref{eq:S'Phi_T2}) is 
\begin{eqnarray}
 S'[\Phi_T]
=\int_0^{\infty}dt\int_0^{\infty}ds\, e^{-t-s}
\frac{(t+s)^2}{12\pi^2}
=\frac{1}{2\pi^2},
\label{eq:evalTac}
\end{eqnarray}
which is the same as a D-brane tension.
Namely, the vacuum energy of $\Phi_T$ is the same as
that of the tachyon vacuum solution \cite{Erler:2007xt, Gorbachev:2010zz}
in the original theory.

\subsection{Half-brane solution}

Let us consider the vacuum energy for $\Phi_H$ (\ref{eq:Phi_H}).
Namely in  (\ref{eq:S'Phif'}), we use
\begin{eqnarray}
&&f'=\frac{1}{1+iG'}=\frac{1}{1+K'}-i\frac{G'}{1+K'}.
\label{eq:half-brane-f}
\end{eqnarray}
Noting the structure of the Chan-Paton factor, we have
\begin{eqnarray}
&&S'[\Phi_H]=\frac{1}{6}(-A_1+A_2),\\
&&A_1=(cG'Bc,\hat Q'(cG'Bc))',~~~
A_2= (cG'BcG',\hat Q'(cG'Bc)G')',
\label{eq:A1A2def}
\end{eqnarray}
where we have used the notation:
\begin{eqnarray}
&&(\Phi,\Psi)'\equiv\langle\!\langle
\Phi\frac{1}{1+K'}\Psi\frac{1}{1+K'}\rangle\!\rangle
=\int_0^{\infty}dt\int_0^{\infty}ds\, e^{-t-s}
\langle\!\langle
\Phi e^{-t K'}\Psi e^{-s K'}\rangle\!\rangle.
\label{eq:marukakko}
\end{eqnarray}
Some relations in worldsheet supersymmetric transformation investigated
in  \cite{Erler:2010pr} hold in the following sense:
\begin{eqnarray}
&&\langle\!\langle G'\cdots \rangle\!\rangle=\frac{1}{2}
\langle\!\langle \delta'(\cdots) \rangle\!\rangle,
~~~\delta'\hat Q'=\hat Q'\delta',
~~~\delta'\partial'=\partial'\delta',\\
&&\delta'(\Phi\Psi)=(\delta'\Phi)\Psi+(-1)^{F(\Phi)}\Phi(\delta'\Psi).
\end{eqnarray}
Using the above, $A_1$ and $A_2$ 
given in  (\ref{eq:A1A2def}) are rewritten as
\begin{eqnarray}
&&A_1=-(\gamma^2,cK')'+5(\gamma^2,c\partial c
 B)'-4(cB\gamma,\gamma K'c)'
\nonumber\\
&&~~~~~~
+2(Bc\gamma,\partial c\gamma)'-2(cB\gamma,\partial\gamma c)'
+2(\gamma,K'\gamma c)',
\label{eq:(1)result}\\
&&A_2=(B\gamma^2,K' c\partial c)'+4(Bc\gamma K',c\gamma K')'
+2(Bc\partial \gamma, c\gamma K')'-2(B\gamma\partial c,c\gamma K')'
\label{eq:(2)result}
\\
&&~~~~~~
+(B\gamma\partial \gamma,c\partial c)'
-(\gamma^2,K' c
 K')'-(cB\gamma,\partial\gamma\partial c)'
-2(cB\gamma,\partial^2\gamma c)'
+(cB\gamma,\gamma\partial^2c)',
\nonumber
\end{eqnarray}
respectively.
To evaluate each term explicitly, following formulas are useful:
\begin{eqnarray}
&&
\langle\!\langle
B ce^{-rK'}ce^{-sK'}\gamma e^{-tK'}\gamma e^{-u K'}
\rangle\!\rangle
=-\frac{r T}{2\pi^2}\cos\frac{\pi t}{T},\\
&&
\langle\!\langle
B ce^{-rK'}\gamma e^{-sK'}c e^{-tK'}\gamma e^{-u K'}
\rangle\!\rangle
=\frac{(r+s)T}{2\pi^2}\cos\frac{\pi (s+t)}{T},\\
&&
\langle\!\langle
B ce^{-rK'}\gamma e^{-sK'}\gamma e^{-tK'}c e^{-u K'}
\rangle\!\rangle
=-\frac{(r+s+t)T}{2\pi^2}\cos\frac{\pi s}{T},
\end{eqnarray}
($T\equiv r+s+t+u$), which are computed as (\ref{eq:evalTacpre}).
We should note that these values are the same as undeformed ones,
namely $K'\to K$
thanks to the identity (\ref{eq:mat=1}) in the deformed matter sector.
Using these formulas, it turns out that
\begin{eqnarray}
 &&A_1=\frac{3}{\pi^2}-\frac{24}{\pi^4},~~
A_2=\frac{9}{2\pi^2}-\frac{24}{\pi^4},~~~~~~~~
S'[\Phi_H]=\frac{1}{4\pi^2}.
\end{eqnarray}
Therefore, the vacuum energy of $\Phi_H$ is a half of that of $\Phi_T$,
namely, $S'[\Phi_H]=\frac{1}{2}S'[\Phi_T]$ from (\ref{eq:evalTac})
and thus we call $\Phi_H$ a ``half-brane'' solution as in
\cite{Erler:2010pr}.

\section{Gauge invariant overlaps for the solutions
\label{sec:overlap}}

In this section, we evaluate gauge invariant overlaps for solutions
$\Phi_T$ (\ref{eq:Phi_T}) and $\Phi_H$ (\ref{eq:Phi_H}) in the theory
with the action (\ref{eq:action'}).
Here, we define a gauge invariant
$\langle\!\langle\Phi\rangle\!\rangle_{\cal V}$
for a string field $\Phi$ after \cite{Erler:2010pr} as
\begin{eqnarray}
&&\langle\!\langle \Phi\rangle\!\rangle_{\cal V}
=\frac{1}{2}{\rm Tr}(\sigma_3\langle I|{\cal V}(i)|\Phi\rangle),
\label{eq:ginvdef}
\end{eqnarray}
where ${\cal V}(i)$ denotes a midpoint insertion of a closed string
vertex operator in the NS-NS sector of the form
$c\bar c \delta(\gamma)\delta(\bar\gamma)V_{\rm m}(z,\bar{z})$
and 
$V_{\rm m}(z,\bar{z})$ is a superconformal matter primary field with
dimension $(1/2,1/2)$.
The above (\ref{eq:ginvdef}) is given by replacing $Y_{-2}$ 
with ${\cal V}(i)$ in (\ref{eq:ll_A_gg}).
Because of the Chan-Paton factor $\sigma_3$, only the GSO$(+)$ sector of 
a string field $\Phi$ gives nontrivial contribution.

In general, for a solution $\Phi_{f'}$
(\ref{eq:Phif'}) to the equation of motion
(\ref{eq:EOM'}), the value of the above gauge invariant overlap is
calculated as
\begin{eqnarray}
&&\langle\!\langle \Phi_{f'}\rangle\!\rangle_{\cal V}
=\langle\!\langle c\frac{K'f'}{1-f'}Bcf'\rangle\!\rangle_{\cal V},
\label{eq:ginv_f'}
\end{eqnarray}
where we have used
\begin{eqnarray}
&&\langle\!\langle\Phi\Psi\rangle\!\rangle_{\cal V}
=(-1)^{(\epsilon(\Phi)+F(\Phi))(\epsilon(\Psi)+F(\Psi))}
\langle\!\langle\Psi\Phi\rangle\!\rangle_{\cal V},
\\
&&\langle\!\langle\hat Q'(\cdots )\rangle\!\rangle_{\cal V}=0.
\label{eq:IcalVQ'}
\end{eqnarray}
The first equation implies the cyclic symmetry and 
the second equation comes from the gauge invariance.

\subsection{Tachyon vacuum solution}

In the case of $\Phi_T$ (\ref{eq:Phi_T}),
that is, the case of $f'=\frac{1}{1+K'}$,
the gauge invariant overlap (\ref{eq:ginv_f'}) is
simplified as
\begin{eqnarray}
&&\langle\!\langle \Phi_T\rangle\!\rangle_{\cal V}
=\langle\!\langle c\frac{1}{1+K'}\rangle\!\rangle_{\cal V}
=\int_0^{\infty}dt\,e^{-t}
\langle\!\langle c e^{-tK'}\rangle\!\rangle_{\cal V}.
\label{eq:PhiTcalV1}
\end{eqnarray}
Using relations
\begin{eqnarray}
&&\frac{1}{2}({\cal L}_0^{\prime}-{\cal
 L}_0^{\prime\dagger})c=-c,
~~~~
\frac{1}{2}({\cal L}_0^{\prime}-{\cal
 L}_0^{\prime\dagger})K'=K',
\label{eq:L0'-L0'dagger1}
\end{eqnarray}
where ${\cal L}'_0=\{Q',{\cal B}_0\}$ and $
{\cal L}_0^{\prime}-{\cal
 L}_0^{\prime\dagger}$
is a derivation with respect to the star product of string fields, we
have
\begin{eqnarray}
&&ce^{-tK'}=t^{1+\frac{1}{2}({\cal L}'_0-{\cal
  L}_0^{\prime\dagger})}(ce^{-K'}).
\label{eq:ce-tK'=}
\end{eqnarray}
In addition, we have the equations
\begin{eqnarray}
\langle I|{\cal V}(i)({\cal B}_0-{\cal B}_0^{\dagger})=0,
~~~\langle I|{\cal V}(i)({\cal L}'_0-{\cal
  L}_0^{\prime\dagger})=0,
\label{eq:IcalV()=0}
\end{eqnarray}
where the second equation is derived from the first equation
(see \cite{Kishimoto:2008zj} for example) and (\ref{eq:IcalVQ'}).
Therefore, (\ref{eq:ce-tK'=})  and
(\ref{eq:IcalV()=0}) imply
\begin{eqnarray}
 \langle\!\langle c e^{-tK'}\rangle\!\rangle_{\cal V}
=t\langle\!\langle c e^{-K'}\rangle\!\rangle_{\cal V},
\end{eqnarray}
and  the integration with respect to $t$
in (\ref{eq:PhiTcalV1}) can be performed explicitly:
\begin{eqnarray}
&&\langle\!\langle \Phi_T\rangle\!\rangle_{\cal V}
=\int_0^{\infty}dt\,e^{-t} t
\langle\!\langle c e^{-K'}\rangle\!\rangle_{\cal V}
=\langle\!\langle c\,e^{-K'}\rangle\!\rangle_{\cal V}.
\label{eq:PhiTcalV2}
\end{eqnarray}
For computation of $e^{-K'}$, we can apply (\ref{eq:e-tKp}) as in the
case of the evaluation of the action:
\begin{eqnarray}
\langle\!\langle \Phi_T\rangle\!\rangle_{\cal V}
&=&\frac{2}{\pi}\langle 0|\tilde{\cal V}(i\infty)\hat{U}_1\,
\tilde c(\frac{\pi}{4})
e^{-\frac{\pi}{2}{\cal C}}{\bf T}\exp\!\left(
\frac{\pi}{4}\int_{-1}^1 du \int_{-\infty}^{\infty}dt'
f_a(t')
\tilde J^a(it'+\frac{\pi}{4}u)\right)\!|0\rangle
\nonumber\\
&=&
\frac{e^{-\frac{\pi}{2}{\cal C}}}{\pi}
\left\langle
\tilde {\cal V}(i\infty)\tilde c(\frac{\pi}{2})
\exp\left(
\int_{-\frac{\pi}{2}}^{\frac{\pi}{2}} du \int_{-\infty}^{\infty}dt'
f_a(t')
\tilde J^a(2it'+u)\right)
\right\rangle.
\label{eq:ginvPhi_Tf}
\end{eqnarray}
Furthermore, using the ghost structure of the closed string vertex
\begin{eqnarray}
 {\cal V}(i)=c(i)c(-i)\delta(\gamma(i))\delta(\gamma(-i))V_{\rm m}(i,-i),
\end{eqnarray}
the above is computed as
\begin{eqnarray}
&&\langle\!\langle \Phi_T\rangle\!\rangle_{\cal V}
=
\frac{e^{-\frac{\pi}{2}{\cal C}}}{\pi}
\lim_{M\to \infty}
\Biggl[
\langle
\delta(\tilde\gamma(iM))\delta(\tilde\gamma(-iM))\rangle_{\beta\gamma}
\langle \tilde c(iM)\tilde c(-iM)\tilde c(\frac{\pi}{2})
\rangle_{bc}
\nonumber\\
&&~~~~~~~~~~~~~~~~~\times
\left\langle\!
\exp\left(
\int_{-\frac{\pi}{2}}^{\frac{\pi}{2}} du \int_{-\infty}^{\infty}dt'
f_a(t')
\tilde J^a(2it'+u)\right)\tilde V_{\rm m}(iM,-iM)\!
\right\rangle_{\rm mat}\Biggr]
\nonumber\\
&&=
\frac{e^{-\frac{\pi}{2}{\cal C}}}{\pi}
\lim_{M\to \infty}
\frac{e^{2M}}{4}\left\langle\!
\exp\left(
\int_{-\frac{\pi}{2}}^{\frac{\pi}{2}} du \int_{-\infty}^{\infty}dt'
f_a(t')
\tilde J^a(2it'+u)\right)\tilde V_{\rm m}(iM,-iM)\!
\right\rangle_{\rm mat}.~~~~~~~~
\label{eq:ginvPhi_Tf2}
\end{eqnarray}
Generally, this value depends on $\tilde J^a$ and $\tilde V_{\rm m}$ in
the matter sector.

In order to perform further explicit calculations,
let us consider the case of (\ref{eq:Vm=e^kX^9}) for
a closed string vertex in the gauge invariant overlap
and $f_a\tilde J^a=f \tilde J=f \frac{i}{\sqrt{2\alpha'}}\tilde \partial
\tilde X^9$, i.e. ${\bm J}(z,\theta)=\psi^9(z) +\theta
\frac{i}{\sqrt{2\alpha'}}\partial X^9(z)$ as a supercurrent,
for the identity-based marginal solution $\Psi_J$
(\ref{eq:PsiJdef}).
We expand the exponential in  (\ref{eq:ginvPhi_Tf2}) as
\begin{eqnarray}
\langle\!\langle \Phi_T\rangle\!\rangle_{\cal V}
&=&\frac{e^{-\frac{\pi}{2}{\cal C}}}{\pi}\sum_{n=0}^{\infty}I^{(k_9)}_n,~~~
\label{eq:exp=sumI^V}
\end{eqnarray}
where $I^{(k_9)}_n$ is the $n$-th order term of $\tilde J$:
\begin{eqnarray}
I^{(k_9)}_n=\frac{1}{n!}\lim_{M\to \infty}
\frac{e^{2M}}{4}\left\langle\!
\left(
\int_{-\frac{\pi}{2}}^{\frac{\pi}{2}} du \int_{-\infty}^{\infty}dt'
f(t')
\tilde J(2it'+u)\right)^n\,
\tilde V_{\rm m}(iM,-iM)
\right\rangle.
\end{eqnarray}
In the case of the lowest order term, $I_0^{(k_9)}$ corresponds to
the undeformed background and it is computed as
\begin{eqnarray}
 I_0^{(k_9)}&=&
\lim_{M\to \infty}
\frac{e^{2M}}{4}\left\langle\!
 e^{\frac{i}{2}k_9\tilde X^9(iM)}e^{-\frac{i}{2}k_9\tilde X^9(-iM)}
\right\rangle\nonumber\\
&=&\lim_{M\to \infty}
\frac{e^{2M}}{4}\frac{1}{\sin(2iM)}=-\frac{i}{2}.
\end{eqnarray}
To evaluate the other terms, following relation among CFT correlators,
which is similar to (\ref{eq:Wardtilde}), is useful:
\begin{eqnarray}
&&\left\langle \tilde J(\tilde z)
\tilde J(\tilde z_1)\cdots \tilde J(\tilde z_n)
\tilde V_{\rm m}(\tilde w,\tilde{\bar{w}})\right\rangle
\nonumber\\
&=&\sum_{i=1}^n
\frac{1}{\sin^2(\tilde z-\tilde z_i)}
\left\langle \tilde J(\tilde z_1)
\cdots \tilde J(\tilde z_{i-1})
\tilde J(\tilde z_{i+1})\cdots
		  \tilde J(\tilde z_n)
\tilde V_{\rm m}(\tilde w,\tilde{\bar{w}})\right\rangle
\nonumber\\
&&+\frac{k_9\sqrt{2\alpha'}}{2\cos\tilde z}
\left(\frac{\cos\tilde w}{\sin(\tilde z-\tilde w)}
-
\frac{\cos\tilde{\bar{w}}}{\sin(\tilde z-\tilde{\bar{w}})}
\right)
\left\langle\tilde J(\tilde z_1)\cdots
 \tilde J(\tilde z_n)
\tilde V_{\rm m}(\tilde w,\tilde{\bar{w}})\right\rangle.
\end{eqnarray}
Then, we have
\begin{eqnarray}
&&I_1^{(k_9)}=\lim_{M\to \infty}
\int_{-\frac{\pi}{2}}^{\frac{\pi}{2}}\!du\int_{-\infty}^{\infty}\!dtf(t)
\frac{e^{2M}}{4}
\left\langle\! \frac{i}{\sqrt{2\alpha'}}\tilde\partial \tilde X^9
(2it +u)\,
 e^{\frac{i}{2}k_9\tilde X^9(iM)}e^{-\frac{i}{2}k_9\tilde X^9(-iM)}
\right\rangle\nonumber\\
&&=\lim_{M\to \infty}\Biggl[
\int_{-\frac{\pi}{2}}^{\frac{\pi}{2}}
\!\!\!du\!
\int_{-\infty}^{\infty}
\!\!\!dtf(t)
\frac{k_9\sqrt{2\alpha'}\cos iM}{2\cos(2it+u)}
\!\left(
\frac{1}{\sin(2it+u-iM)}
-
\frac{1}{\sin(2it+u+iM)}
\right)
\nonumber\\
&&~~~~~~~~~~~~~\times
\frac{e^{2M}}{4}
\left\langle\! 
 e^{\frac{i}{2}k_9\tilde X^9(iM)}e^{-\frac{i}{2}k_9\tilde X^9(-iM)}
\right\rangle
\Biggr]
\nonumber\\
&&=i\pi k_9\sqrt{2\alpha'}\int_{-\infty}^{\infty}
\!\!\!dtf(t)\,I_0^{(k_9)}
\equiv \hat I_1^{(k_9)}\,I_0^{(k_9)},
\end{eqnarray}
for the first order and
\begin{eqnarray}
I_2^{(k_9)}
&=&\frac{1}{2}
\int_{-\frac{\pi}{2}}^{\frac{\pi}{2}}
\!\!\!du_1\!
\int_{-\infty}^{\infty}
\!\!\!dt_1f(t_1)
\int_{-\frac{\pi}{2}}^{\frac{\pi}{2}}
\!\!\!du_2\!
\int_{-\infty}^{\infty}
\!\!\!dt_2f(t_2)
\Biggl[
\frac{1}{\sin^2(u_1-u_2+2i(t_1-t_2))}
I_0^{(k_9)}
\nonumber\\
&&+\lim_{M\to \infty}\Biggl\{
\frac{k_9\sqrt{2\alpha'}\cos iM}{2\cos(2it_1+u_1)}
\!\left(
\frac{1}{\sin(2it_1+u_1-iM)}
-
\frac{1}{\sin(2it_1+u_1+iM)}
\right)
\nonumber\\
&&~~~~~~~~~
\times
\frac{e^{2M}}{4}
\left\langle\! \frac{i}{\sqrt{2\alpha'}}\tilde\partial \tilde X^9
(2it_2 +u_2)\,
 e^{\frac{i}{2}k_9\tilde X^9(iM)}e^{-\frac{i}{2}k_9\tilde X^9(-iM)}
\right\rangle
\Biggr\}\Biggr]
\nonumber\\
&=&I_2I_0^{(k_9)}+\frac{1}{2}\hat{I}_1^{(k_9)}I_1^{(k_9)}
=\left(I_2+\frac{1}{2}(\hat{I}_1^{(k_9)})^2\right)I_0^{(k_9)}
\end{eqnarray}
for the second order,
where we have used (\ref{eq:deltafn}) and
\begin{eqnarray}
&&I_2=\frac{\pi^2}{2}\int_{-\infty}^{\infty}dt(f(t))^2=\frac{\pi}{2}{\cal
 C}.
\end{eqnarray}
((\ref{eq:calCdef}) with $\Omega=2$, where $X^9(y)X^9(z)\sim
-2\alpha'\log(y-z)$ in our convention, is used for the last equality.)
In the same way, higher order terms can be computed and the results for
even order terms and odd order terms are given by
\begin{eqnarray}
I_{2n}^{(k_9)}&=&I_0^{(k_9)}\sum_{l=0}^n\frac{1}{(n-l)!\,(2l)!}I_2^{n-l}
(\hat{I}_1^{(k_9)})^{2l},\\
I_{2n-1}^{(k_9)}&=&I_0^{(k_9)}\sum_{l=0}^{n-1}\frac{1}{(n-1-l)!\,(2l+1)!}
I_2^{n-1-l}(\hat{I}_1^{(k_9)})^{2l+1},
\end{eqnarray}
respectively.
Thus the gauge invariant overlap is calculated as
\begin{eqnarray}
\langle\!\langle \Phi_T\rangle\!\rangle_{\cal V}
=
\frac{e^{-\frac{\pi}{2}{\cal C}}}{\pi}
\sum_{n=0}^{\infty}I_{n}^{(k_9)}
=\frac{e^{-\frac{\pi}{2}{\cal
C}}}{\pi}e^{I_2+\hat{I}_1^{(k_9)}}I_0^{(k_9)}
=\frac{1}{\pi}e^{\hat{I}_1^{(k_9)}}I_0^{(k_9)}
=\frac{1}{2\pi i}\,e^{\hat{I}_1^{(k_9)}}.
\label{eq:PhiTVresult}
\end{eqnarray}
The exponent of the above can be rewritten as
\begin{eqnarray}
 \hat{I}_1^{(k_9)}&=&i\pi
  k_9\sqrt{2\alpha'}\int_{-\infty}^{\infty}dtf(t)
=i\pi k_9\sqrt{\alpha'}\int_{C_{\rm L}}\frac{dz}{2\pi i}
F(z).
\end{eqnarray}
Therefore, the phase factor $e^{\hat{I}_1^{(k_9)}}$, which is induced
by a current $J=\frac{i}{\sqrt{2\alpha'}}\partial X^9$, is exactly the
same as that in (\ref{eq:ginv_phase}) obtained by a different method.
It corresponds to the phase factor appeared in \cite{Inatomi:2012nv,
Katsumata:2004cc} in the case of bosonic SFT.

\subsection{Half-brane solution}

In the case of $\Phi_H$ (\ref{eq:Phi_H}), or $f'$ given in
(\ref{eq:half-brane-f}), the gauge invariant overlap
(\ref{eq:ginv_f'}) is simplified as
\begin{eqnarray}
 \langle\!\langle \Phi_H\rangle\!\rangle_{\cal V}
=-\langle\!\langle cG' Bc \frac{G'}{1+K'}\rangle\!\rangle_{\cal V}
=-\int_0^{\infty}dt e^{-t}
\langle\!\langle cG' Bc G'e^{-tK'}\rangle\!\rangle_{\cal V},
\label{eq:ginvPhi_H1}
\end{eqnarray}
thanks to the structure of Chan-Paton factor.
Using (\ref{eq:L0'-L0'dagger1}), (\ref{eq:IcalV()=0}) and
\begin{eqnarray}
&&\frac{1}{2}({\cal L}_0^{\prime}-{\cal
 L}_0^{\prime\dagger})B=B,
~~~~\frac{1}{2}({\cal L}_0^{\prime}-{\cal
 L}_0^{\prime\dagger})G'=\frac{1}{2}G',\\
&&t^{\frac{1}{2}({\cal L}_0^{\prime}-{\cal
 L}_0^{\prime\dagger})}(cG'BcG'e^{-K'})
= cG' Bc G'e^{-tK'},\\
&&-cK'e^{-K'}=
\frac{1}{2}({\cal L}_0^{\prime}-{\cal
 L}_0^{\prime\dagger})(ce^{-K'})+ce^{-K'},
\end{eqnarray}
which are a deformed version of relations in \cite{Erler:2010pr},
the integration with respect to $t$ in (\ref{eq:ginvPhi_H1})
can be explicitly performed and (\ref{eq:ginvPhi_H1}) is simplified as
\begin{eqnarray}
\langle\!\langle \Phi_H\rangle\!\rangle_{\cal V}
&=&-\langle\!\langle cG' Bc G'e^{-K'}\rangle\!\rangle_{\cal V}
=-\langle\!\langle cB(cK'+2i\gamma G')e^{-K'}\rangle\!\rangle_{\cal V}
\nonumber\\
&=&-\langle\!\langle (cK'+2icB\gamma G')e^{-K'}\rangle\!\rangle_{\cal V}
=\langle\!\langle (c+2ic\gamma BG')e^{-K'}\rangle\!\rangle_{\cal V}.
\label{eq:ginvPhi_H_temp}
\end{eqnarray}
Furthermore, noting
\begin{eqnarray}
\frac{1}{2}({\cal B}_0-{\cal B}_0^{\dagger})K'=B,~~
\frac{1}{2}({\cal B}_0-{\cal B}_0^{\dagger})B=0,~~
\frac{1}{2}({\cal B}_0-{\cal B}_0^{\dagger})c=0,~~
\frac{1}{2}({\cal B}_0-{\cal B}_0^{\dagger})\gamma=0,
\end{eqnarray}
for a derivation ${\cal B}_0-{\cal B}_0^{\dagger}$ with respect to
the star product,
the second term of the last expression in (\ref{eq:ginvPhi_H_temp})
is calculated as
\begin{eqnarray}
&&2i\langle\!\langle c\gamma BG'e^{-K'}\rangle\!\rangle_{\cal V}
=2i\langle\!\langle G' c\gamma B e^{-K'}\rangle\!\rangle_{\cal V}
=i\langle\!\langle \delta'(c\gamma B e^{-K'})\rangle\!\rangle_{\cal V}
\nonumber\\
&&=\langle\!\langle -2\gamma^2 BG'e^{-K'}\!
+\frac{1}{2}c\partial c B e^{-K'}
\rangle\!\rangle_{\cal V}
=\langle\!\langle 
({\cal B}_0-{\cal B}_0^{\dagger})\!\left(\!
\gamma^2 e^{-K'}-\frac{1}{4}cK'ce^{-K'}
\!\right)\!
-\frac{1}{2}c e^{-K'}
\rangle\!\rangle_{\cal V}
\nonumber\\
&&=-\frac{1}{2}
\langle\!\langle c e^{-K'}
\rangle\!\rangle_{\cal V}.
\end{eqnarray}
Therefore, (\ref{eq:ginvPhi_H_temp}) can be rewritten as
\begin{eqnarray}
 \langle\!\langle \Phi_H\rangle\!\rangle_{\cal V}
&=&
\frac{1}{2}
\langle\!\langle c e^{-K'}
\rangle\!\rangle_{\cal V}
\nonumber\\
&=&\frac{e^{-\frac{\pi}{2}{\cal C}}}{2\pi}
\left\langle
\tilde {\cal V}(i\infty)\tilde c(\frac{\pi}{2})
\exp\left(
\int_{-\frac{\pi}{2}}^{\frac{\pi}{2}} du \int_{-\infty}^{\infty}dt'
f_a(t')
\tilde J^a(2it'+u)\right)
\right\rangle.
\label{eq:ginvK'1}
\end{eqnarray}
This value is just a half of  (\ref{eq:ginvPhi_Tf}),
namely, $\langle\!\langle \Phi_H\rangle\!\rangle_{\cal V}
=\frac{1}{2}\langle\!\langle \Phi_T\rangle\!\rangle_{\cal V}$.
We notice that this relation itself does not depend on 
details of $\tilde J^a$ and $\tilde V_{\rm m }$ in the matter sector.

\section{Concluding remarks
\label{sec:remarks}}

We have considered a version of ``$GKBc\gamma$ algebra'' in the theory
around an identity-based marginal solution $\Psi_J$, which is made of
supercurrents associated with a Lie algebra in the matter sector,
in modified cubic SSFT.
Corresponding to a deformed BRST operator $Q'$, string fields $G$ and
$K$ are deformed to $G'$ and $K'$, respectively.
Using these ingredients, we constructed a tachyon vacuum 
solution $\Phi_T$ and a half-brane solution $\Phi_H$
 and evaluated the vacuum energy and gauge invariant overlap
for them.
The values of the vacuum energy for these solutions are 
exactly the same as those of the tachyon vacuum solution $\Psi_T$ and
the half-brane solution $\Psi_H$ in the original theory, respectively.
Namely, $S'[\Phi_T]=S[\Psi_T]=2S[\Psi_H]=2S'[\Phi_H]$ holds, where we
note the relation: $S'[\Phi]=S[\Phi+\Psi_J\sigma_3]-S[\Psi_J\sigma_3]$
(\ref{eq:action'}) in general.
By introducing a parameter $s$ in the weighting function:
$F_a(z)\to s F_a(z)$ and taking a differentiation of the action 
 and integration from $s=0$ to $s=1$, 
we can show $S[\Phi_T+\Psi_J\sigma_3]=S[\Psi_T]$ and 
$S[\Phi_H+\Psi_J\sigma_3]=S[\Psi_H]$ in the same way as in \cite{Inatomi:2012nv}.
Therefore, from consistency, 
the vacuum energy of $\Psi_J$ vanishes: $S[\Psi_J\sigma_3]=0$
although direct computation of $S[\Psi_J\sigma_3]$ is difficult due to
singular property of the identity state.

The values of the gauge invariant overlap for both of $\Phi_T$ and
$\Phi_H$ change according to the marginal current in $\Psi_J$.
However, the relation between them:
 $\langle\!\langle \Phi_H\rangle\!\rangle_{\cal V}
=\frac{1}{2}\langle\!\langle \Phi_T\rangle\!\rangle_{\cal V}$ 
holds as in the case of the original theory.
If we take a closed tachyon vertex for a Dirichlet direction as
the matter part of ${\cal V}$ and $\partial X^9$ as a current $J$,
a phase factor appears in the gauge invariant overlap
and it is consistent with the effect caused by a field redefinition
induced by $\Psi_J$.

These results in the above support the expectation that 
the identity-based solution $\Psi_J$ corresponds to a marginal
deformation
and they are an extension of results in \cite{Inatomi:2012nv} for
bosonic SFT to modified cubic SSFT.

If we take a zero momentum graviton vertex $\psi^{\mu}\bar\psi^{\nu}$
as a matter part of closed string vertex
and $J\sim \partial X^9$ in the gauge invariant overlap,
the values for $\Phi_T$ and $\Phi_H$ are exactly the same as 
$\Psi_T$ and $\Psi_H$
in the undeformed background, respectively (i.e. 
$\langle\!\langle \Phi_T\rangle\!\rangle_{\cal V}
=\langle\!\langle \Psi_T\rangle\!\rangle_{\cal V}
=2\langle\!\langle \Psi_H\rangle\!\rangle_{\cal V}
=2\langle\!\langle \Phi_H\rangle\!\rangle_{\cal V}$),
which can be proved in the same way as the evaluation of the action thanks
to (\ref{eq:mat=1}).
These are reminiscent of the relation between the energy and the gauge
invariant overlap in bosonic SFT \cite{Baba:2012cs}.

In this paper, we considered the NS sector without GSO projection and as
a closed string vertex for the gauge invariant overlap, we have only
considered the NS-NS sector.
It may be interesting to investigate gauge invariant overlaps
for closed string vertices in the R-R sector.

Here, we have considered a theory only around an identity-based
marginal solution in SSFT and we found that $G'K'Bc\gamma$ algebra
has the same algebraic structure with undeformed $GKBc\gamma$ algebra.
If we consider a theory around another type of identity-based 
universal solutions found in \cite{Inatomi:2011an},
we expect that the algebraic structure might be changed when a homotopy
operator exists.
Using such an algebra, vacuum energy and/or gauge invariant overlap
might be evaluated directly.

\section*{Acknowledgments}

The work of I.~K. and T.~T. is supported by
JSPS Grant-in-Aid for Scientific Research (B) (\#24340051).
The work of I.~K. is supported partly by 
Grant for Promotion of Niigata University Research Projects
and 
partly by
Grant-in-Aid for Research Project from Institute of Humanities, Social
Sciences and Education, Niigata University.

\appendix

\section{A brief review of $KBc$ algebra and its extension
\label{sec:convention}}
Here, we summarize some results on a supersymmetric extension of 
 $KBc$ algebra developed in \cite{Erler:2010pr}
and we list our convention and notation in this paper.
We consider string fields in the NS sector without GSO projection and
 therefore we introduce Chan-Paton factors, which are represented by
 $2\times 2$  Pauli matrices: $\sigma_i~(i=1,2,3)$ (and the identity matrix 
implicitly).
There are four sectors corresponding to
Grassmann parity ($\epsilon$) and worldsheet spinor ($F$)
and we assign Chan-Paton factors as in Table \ref{tab:CPmat}.
\begin{table}[h]
\begin{center}
 \begin{tabular}{|c|c|c|c|c|c|}
\hline
Grassmann parity ($\epsilon$)&worldsheet spinor ($F$)& Chan-Paton factor\\
\hline
\hline
even&even&$1$\\
\hline
odd&even&$\sigma_3$\\
\hline
even&odd&$\sigma_2$\\
\hline
odd&odd&$\sigma_1$\\
\hline
\end{tabular}
\end{center}
\caption{Assignment of the Chan-Paton factor
\label{tab:CPmat}
}
\end{table}
The NS action $S[\Psi]$ is written as
\begin{eqnarray}
S[\Psi]&=&\frac{1}{2}\langle\!\langle
\Psi\,\hat Q\Psi\rangle\!\rangle
+\frac{1}{3}\langle\!\langle\Psi^3\rangle\!\rangle,
\label{eq:action}
\end{eqnarray}
where we omit a symbol for the star product among string fields and 
$\langle\!\langle \cdot \rangle\!\rangle$ includes a trace for $2\times
2$ Chan-Paton matrices and a picture changing operator with picture number
$(-2)$ denoted by $Y_{-2}$:
\begin{eqnarray}
\langle\!\langle A\rangle\!\rangle 
&\equiv& \frac{1}{2}{\rm Tr}
\left(\sigma_3\langle I|Y_{-2} A\rangle\right).
\label{eq:ll_A_gg}
\end{eqnarray}
Here, $\langle I|$ is the identity state and we denote $\hat Q\equiv
Q_{\rm B}\sigma_3$ in (\ref{eq:action}). 
The NS string field $\Psi$ in the action  (\ref{eq:action}) can be expanded as
\begin{eqnarray}
 \Psi&=&\Psi_{+}\sigma_3+\Psi_{-}\sigma_2,
\label{eq:Psi_++Psi_-}
\end{eqnarray}
where $\Psi_{+}$ ($\Psi_{-}$) is in the GSO$(+)$ (GSO$(-)$) sector.
In general, we have
\begin{eqnarray}
&&\hat Q(\Phi\Psi)=(\hat
 Q\Phi)\Psi+(-1)^{\epsilon(\Phi)+F(\Phi)}\Phi(\hat Q\Psi),
\label{eq:hatQPhiPsi}\\
&&\langle\!\langle\Phi\Psi\rangle\!\rangle
=(-1)^{(\epsilon(\Phi)+F(\Phi))(\epsilon(\Psi)+F(\Psi))}
\langle\!\langle\Psi\Phi\rangle\!\rangle,
\\
&&\langle\!\langle\hat Q(\cdots )\rangle\!\rangle=0.
\label{eq:hatQ()}
\end{eqnarray}
In this paper, we define $Y_{-2}$ in (\ref{eq:ll_A_gg}) 
using two inverse picture changing operators:
\begin{eqnarray}
&&Y_{-2}=Y(i)Y(-i),~~~~~Y(z)\equiv c(z)\delta'(\gamma(z)).
\label{eq:Y_-2def}
\end{eqnarray}
In the above convention, we define string fields $K,B,c$ as
\begin{eqnarray}
&&K=\frac{\pi}{2}K_1^LI,~~~~B=\frac{\pi}{2}B_1^LI\sigma_3,
~~~~c=\frac{1}{\pi}c(1)I\sigma_3
=\frac{2}{\pi}\hat{U}_1\tilde c(0)|0\rangle\sigma_3.
\label{eq:KBcdef}
\end{eqnarray}
In the above, $K_1^L,B_1^L,\hat U_1$ are defined as in
\cite{Schnabl:2005gv}:
\begin{eqnarray}
\label{eq:K1Ldef}
&&K_1^L=\{Q_{\rm B},B_1^L\},
~~B_1^L=\frac{1}{2}B_1+\frac{1}{\pi}({\cal B}_0+{\cal
B}_0^{\dagger}),
~~\hat{U}_r=U_r^{\dagger}U_r,~~U_r=\left(\frac{2}{r}\right)^{{\cal
L}_0},
\\
&&B_1=b_{-1}+b_1,~~~{\cal
 B}_0=b_0+\sum_{k=1}^{\infty}\frac{2(-1)^{k+1}}{4k^2-1}b_{2k},
~~~{\cal L}_0=\{Q_{\rm B},{\cal B}_0\},
\end{eqnarray}
and $\tilde c(\tilde z)=(\cos\tilde z)^2\,c(\tan \tilde z)$
is $c$-ghost in the sliver frame.

Similarly, we define $G,\gamma$ in the case of superstring as
\cite{Erler:2010pr}:
\begin{eqnarray}
&&G={\cal G}_LI\sigma_1,~~~~
\gamma=\frac{1}{\sqrt{\pi}}\gamma(1)I\sigma_2
=\sqrt{\frac{2}{\pi}}\hat{U}_1\tilde \gamma(0)|0\rangle\sigma_2,
\label{eq:Ggammadef}
\end{eqnarray}
where $\tilde \gamma(\tilde z)=(\cos\tilde z)\,\gamma(\tan\tilde z)$ 
is $\gamma$-ghost in the sliver frame and
\begin{eqnarray}
\label{eq:calG_L}
&&{\cal G}_L=\frac{1}{2}({\cal G}+{\cal G}^{\star}),\\
&&{\cal G}=\oint \frac{dz}{2\pi i}\sqrt{\frac{\pi}{2}}\sqrt{1+z^2}G(z)
=\sqrt{\frac{\pi}{2}}\sum_{n=0}^{\infty}
\begin{pmatrix}
1/2\\
n
\end{pmatrix}
G_{2n-\frac{1}{2}},\\
&&{\cal G}^{\star}=\oint \frac{dz}{2\pi i}\sqrt{\frac{\pi}{2}}z\sqrt{1+z^{-2}}G(z)
=\sqrt{\frac{\pi}{2}}\sum_{n=0}^{\infty}
\begin{pmatrix}
1/2\\
n
\end{pmatrix}
G_{\frac{1}{2}-2n}.
\end{eqnarray}
Here, 
$\begin{pmatrix}
r\\
s
\end{pmatrix}
\equiv \frac{\Gamma(r+1)}{\Gamma(s+1)\gamma(r-s+1)}$
is a binomial coefficient and $G_r=[Q_{\rm B},\beta_r]$ is a superconformal
generator. Actually, ${\cal G}$ can be rewritten as
\begin{eqnarray}
&&{\cal G}=\sqrt{\frac{\pi}{2}}\,[Q_{\rm
 B},\tilde\beta_{-\frac{1}{2}}],
\label{eq:G=Qbeta}
\end{eqnarray}
where $\tilde\beta_{-\frac{1}{2}}$ is a mode of $\tilde \beta(\tilde
z)=(\cos\tilde z)^{-3}\,\beta(\tan\tilde z)$ 
in the sliver frame:
\begin{eqnarray}
 \tilde\beta_{-\frac{1}{2}}=\oint \frac{dz}{2\pi i}\sqrt{1+z^2}\,\beta(z)
=\sum_{n=0}^{\infty}
\begin{pmatrix}
1/2\\
n
\end{pmatrix}
\beta_{2n-\frac{1}{2}}.
\end{eqnarray}
{}From explicit computation using mode expansions, we find 
that ${\cal G},{\cal B}_0$ and $\tilde \beta_{-\frac{1}{2}}$
satisfy
\begin{eqnarray}
&&\{{\cal G},{\cal B}_0\}=\frac{1}{2}\sqrt{\frac{\pi}{2}}\,\tilde
  \beta_{-\frac{1}{2}},~~~~
\{{\cal G},{\cal B}_0^{\dagger}\}
=-\frac{1}{2}\sqrt{\frac{\pi}{2}}\,\tilde
  \beta_{-\frac{1}{2}}.
\label{eq:GBbeta_rel}
\end{eqnarray}
Then, noting $[{\cal G},{\cal L}_0^{\dagger}]=\frac{1}{2}{\cal G}$,
$Q_{\rm
B}|I\rangle=0$ and ${\cal B}_0|I\rangle={\cal B}_0^{\dagger}|I\rangle$,
we have
\begin{eqnarray}
&&{\cal G}|I\rangle=2^{{\cal L}_0^{\dagger}+\frac{1}{2}}{\cal G}|0\rangle=0,\\
&&\tilde \beta_{-\frac{1}{2}}|I\rangle=\sqrt{\frac{2}{\pi}}\{{\cal
 G},{\cal B}_0-{\cal B}_0^{\dagger}\}|I\rangle=0.
\label{eq:betaI=0}
\end{eqnarray}

Among the above string fields, i.e., $G,K,B,c$ and $\gamma$, 
we have following relations:
\begin{eqnarray}
&&Bc+cB=1,~~~G^2=K,~~~~B^2=0,~~~c^2=0,\\
&&\hat QB=K,~~~~\hat QK=0,~~~\hat QG=0,~~~~\hat Qc=cKc-\gamma^2,\\
&&BG=GB,~~BK=KB,~~GK=KG,~~
B\gamma+\gamma B=0,~~c\gamma+\gamma c=0.~~~~
\end{eqnarray}
Furthermore, worldsheet supersymmetry transformations of string fields
are given by
\begin{eqnarray}
 \delta c=2i\gamma,~~~~\delta \gamma=-\frac{i}{2}\partial c,~~~~
\delta G=2K,~~~\delta K=0,~~~\delta B=0,
\end{eqnarray}
where $\partial$ and $\delta$ are defined by
\begin{eqnarray}
 \partial \Phi\equiv K\Phi-\Phi K=\frac{\pi}{2}K_1\Phi,~~~~
\delta \Phi\equiv G\Phi-(-)^{F(\Phi)}\Phi G=({\cal G}\sigma_1)\Phi,
\label{eq:partialdel}
\end{eqnarray}
which satisfy the relation $\delta^2=\partial$.

\section{An expansion formula
\label{sec:expansion}}

We consider a generalization of a well-known formula,
\begin{eqnarray}
 \delta(e^X)=\int_0^1 d\alpha\, e^{(1-\alpha)X}(\delta X)e^{\alpha X}.
\label{eq:Feynman}
\end{eqnarray}
Namely, by expanding as
\begin{eqnarray}
 e^{X+\delta X}=
e^X+\sum_{N=1}^{\infty}(e^{X+\delta X})|_{O((\delta
  X)^N)}
\end{eqnarray}
for $[X,\delta X]\ne 0$ in general, we will find a similar expression as
(\ref{eq:Feynman}) for the order $(\delta X)^N$ term: $(e^{X+\delta
X})|_{O((\delta
  X)^N)}$.
By a straightforward expansion, we have
\begin{eqnarray}
 &&(e^{X+\delta X})|_{O((\delta
  X)^N)}=\sum_{n=0}^{\infty}\frac{1}{n!}(X+\delta X)^n|_{O((\delta
  X)^N)}\nonumber\\
&&=\sum_{n=N}^{\infty}\frac{1}{n!}\sum_{k_1=0}^{n-N}
\sum_{k_2=0}^{n-N-k_1}
\cdots\!\!\!
\sum_{k_N=0}^{n-N-k_1-k_2\cdots -k_{N-1}}\!\!\!\!
X^{k_1}(\delta X)X^{k_2}(\delta X) \cdots X^{k_N}(\delta X)
X^{n-N-k_1-k_2\cdots -k_N}
\nonumber\\
&&=\sum_{k_0=0}^{\infty}
\sum_{k_1=0}^{\infty}\cdots 
\sum_{k_N=0}^{\infty}\frac{1}{(k_0+k_1\cdots +k_N+N)!}
X^{k_0}(\delta X)X^{k_1}(\delta X)\cdots X^{k_{N-1}}(\delta X)
X^{k_N}.
\end{eqnarray}
Noting 
$B(p,q)=\int_0^1dt(1-t)^{p-1}t^{q-1}=\frac{\Gamma(p)\Gamma(q)}{\Gamma(p+q)}
$ and comparing coefficients, we find
 that the above can be rewritten using integrals:
\begin{eqnarray}
  &&(e^{X+\delta X})|_{O((\delta
  X)^N)}\nonumber\\
&&=\int_0^1\!\!du_1\!\int_0^{1-u_1}\!\!\!\!\!\!du_2\cdots \!\!
\int_0^{1-u_1-u_2\cdots -u_{N-1}}\!\!\!\!\!\!\!\!\!du_N\,
e^{(1-u_1-u_2\cdots -u_N)X}(\delta X)e^{u_1 X}(\delta X)
e^{u_2 X}\!\cdots (\delta X) e^{u_N X}.
\nonumber\\
\label{eq:eXdelXN}
\end{eqnarray}

\section{ $J$-independence in the matter sector
\label{sec:Jindep}}

In this section, we demonstrate a formula
\begin{eqnarray}
 e^{-T\frac{\pi}{2}{\cal C}}
\left\langle \exp\left(
\int_{-\frac{\pi}{2}}^{\frac{\pi}{2}}\!\!du
\int_{-\infty}^{\infty}\!\!\!dt' f_a(t')
\tilde J^a(\frac{2it'}{T}\!+\!u)\right)\right\rangle_{\rm mat}
&=&1
\label{eq:mat=1}
\end{eqnarray}
which is essential to evaluate the vacuum energy
for $\Phi_T$ (\ref{eq:Phi_T}) and $\Phi_H$ (\ref{eq:Phi_H}).
The above result was also used in \cite{Inatomi:2012nv}
in the context of bosonic SFT.

Let us define $I_n$ as the order of $J^n$ in the above $\langle \cdots
\rangle$, namely,
\begin{eqnarray}
&&I_n\equiv 
\frac{1}{n!}\!\int_{-\frac{\pi}{2}}^{\frac{\pi}{2}}\!\!du_1\!\!
\int_{-\infty}^{\infty}\!\!\!dt_1f_{a_1}(t_1)
\cdots \!\!
\int_{-\frac{\pi}{2}}^{\frac{\pi}{2}}\!\!du_n\!\!
\int_{-\infty}^{\infty}\!\!\!dt_nf_{a_n}(t_n)
\!\left\langle\!
\tilde J^{a_1}(\frac{2it_1}{T}\!+\!u_1)
\!\cdots\!
\tilde J^{a_n}(\frac{2it_n}{T}\!+\!u_n)\!
\right\rangle\!.
\nonumber\\
\end{eqnarray}
To evaluate $I_n$, we note the relation among CFT correlators:
\begin{eqnarray}
&&\left\langle J^a(z)J^{a_1}(z_1)\cdots J^{a_n}(z_n)\right\rangle
\nonumber\\
&=&\sum_{i=1}^n\Biggl[
\frac{\frac{1}{2}\Omega^{aa_i}}{(z-z_i)^2}
\left\langle J^{a_1}(z_1)\cdots J^{a_{i-1}}(z_{i-1})
J^{a_{i+1}}(z_{i+1})\cdots
		  J^{a_n}(z_n)\right\rangle
\nonumber\\
&&+\frac{f^{aa_i}_{~~~b}}{z-z_i}
\left\langle J^b(z_i)J^{a_1}(z_1)\cdots J^{a_{i-1}}(z_{i-1})
J^{a_{i+1}}(z_{i+1})\cdots
		  J^{a_n}(z_n)\right\rangle
\Biggr],
\end{eqnarray}
which is the Ward identity derived from the OPE
(\ref{eq:JJ_OPE}).\footnote{
It can be obtained by calculating the contour integral
$\oint\frac{dz}{2\pi i}\epsilon_a(z)J^a(z)$ 
inserted in a correlation function of
$J^{a_i}(z_i)$'s. (See \cite{Ginsparg:1988ui} for example.)}
In terms of the sliver frame, we have
\begin{eqnarray}
&&\left\langle \tilde J^a(\tilde z)
\tilde J^{a_1}(\tilde z_1)\cdots \tilde J^{a_n}(\tilde z_n)\right\rangle
\nonumber\\
&=&\sum_{i=1}^n\Biggl[
\frac{1}{\sin^2(\tilde z-\tilde z_i)}\frac{1}{2}\Omega^{aa_i}
\left\langle \tilde J^{a_1}(\tilde z_1)
\cdots \tilde J^{a_{i-1}}(\tilde z_{i-1})
\tilde J^{a_{i+1}}(\tilde z_{i+1})\cdots
		  \tilde J^{a_n}(\tilde z_n)\right\rangle
\nonumber\\
&&+\frac{\cos\tilde z_i}{\cos\tilde z}
\frac{1}{\sin(\tilde z-\tilde z_i)}f^{aa_i}_{~~~b}
\left\langle \tilde J^b(\tilde z_i)
\tilde J^{a_1}(\tilde z_1)\cdots
\tilde J^{a_{i-1}}(\tilde z_{i-1})
\tilde J^{a_{i+1}}(\tilde z_{i+1})\cdots
		  \tilde J^{a_n}(\tilde z_n)\right\rangle
\Biggr].~~~~~~~
\label{eq:Wardtilde}
\end{eqnarray}
Using the above identity and
some relations in (\ref{eq:f_Rel1}), (\ref{eq:f_Rel2}),
we have
\begin{eqnarray}
  I_0=1,~~~I_1=0,~~~~I_2=\frac{\pi}{2}{\cal C}T,~~~~I_3=0,
\label{eq:I0123}
\end{eqnarray}
where ${\cal C}$ is defined in (\ref{eq:calCdef}) and 
we have used \cite{Inatomi:2012nv}:
\begin{eqnarray}
 \int_{-\frac{\pi}{2}}^{\frac{\pi}{2}}\frac{du_1}{\pi}
\int_{-\frac{\pi}{2}}^{\frac{\pi}{2}}\frac{du_2}{\pi}
\frac{1}{
\sin^2(u_1-u_2+2i(t_1-t_2))}
=\delta(t_1-t_2).
\label{eq:deltafn}
\end{eqnarray}
In order to evaluate $I_n$ in general, we define
\begin{eqnarray}
 I_n^{a_1}(z)&\equiv&\frac{1}{n!}\int_{-\frac{\pi}{2}}^{\frac{\pi}{2}}\!du_2\!
\int_{-\infty}^{\infty}\!dt_2f_{a_2}(t_2)
\cdots \!
\int_{-\frac{\pi}{2}}^{\frac{\pi}{2}}\!du_n\!
\int_{-\infty}^{\infty}\!dt_nf_{a_n}(t_n)
\nonumber\\
&&~~~~~~~~~~\times \left\langle\tilde J^{a_1}(z)
\tilde J^{a_2}(\frac{2it_2}{T}+u_2)
\cdots
\tilde J^{a_n}(\frac{2it_n}{T}+u_n)
\right\rangle,~~~~~(n\ge 1)
\end{eqnarray}
so that
\begin{eqnarray}
I_n&=&
\int_{-\frac{\pi}{2}}^{\frac{\pi}{2}}\!du\!
\int_{-\infty}^{\infty}\!dtf_{a_1}(t)
I_n^{a_1}(\frac{2it}{T}+u).
\end{eqnarray}
Then, we have
\begin{eqnarray}
&&I_1^a(z)=0,~~~~
I_2^a(\frac{2it}{T}+u)=\frac{\pi}{4}T\,\Omega^{ab}f_b(t).
\end{eqnarray}
Using mathematical induction with relations in
  (\ref{eq:Wardtilde}),  (\ref{eq:f_Rel1}) and (\ref{eq:f_Rel2}), we can
  show
\begin{eqnarray}
 I_n^a(\frac{2it}{T}+u)=\frac{2}{n}I_2^a(\frac{2it}{T}+u)I_{n-2}
=\frac{1}{n}\frac{\pi}{2}T\,\Omega^{ab}f_b(t)I_{n-2}.
\label{eq:Inaformula}
\end{eqnarray}
This implies $I_n=\frac{2}{n}I_{2}I_{n-2}$ and hence 
\begin{eqnarray}
 I_{2m}=\frac{2^m}{(2m)!!}I_2^mI_0=\frac{1
}{m!}I_2^m,~~~~I_{2m+1}=0,~~~(m=0,1,2,3,\cdots)
\end{eqnarray}
{}from (\ref{eq:I0123}).
Using the above results, we have
\begin{eqnarray}
\sum_{n=0}^{\infty}I_n=\sum_{m=0}^{\infty}
\frac{1}{m!}I_2^m=\exp\left(\frac{\pi}{2}T{\cal C}\right),
\label{eq:kekka}
\end{eqnarray}
which is equivalent to (\ref{eq:mat=1}).

\section{On the gauge invariant overlap and field redefinition
\label{sec:fieldredefinition}}

In this section, we comment on a relation among the gauge invariant
overlaps for string fields related by a field redefinition,
induced by the solution $\Psi_J$ (\ref{eq:PsiJdef}).
In the following, we consider $u(1)$ supercurrent
in the 9-th spatial direction:
${\bm J}(z,\theta)=\psi^9(z) +\theta
\frac{i}{\sqrt{2\alpha'}}\partial X^9(z)$
for the identity-based marginal solution $\Psi_J$ for simplicity.
Then, the BRST operator $Q'$ (\ref{eq:Q'def})
at the solution $\Psi_J$ (\ref{eq:PsiJdef}) 
can be rewritten as a similarity transform from the conventional 
BRST operator $Q_{\rm B}$ as \cite{Kishimoto:2005bs}:
\begin{eqnarray}
 Q'=e^{\frac{i}{2\sqrt{\alpha'}}X(F)}Q_{\rm B}
e^{-\frac{i}{2\sqrt{\alpha'}}X(F)},
\end{eqnarray}
where $X(F)$ is given by an integration along a unit circle:
\begin{eqnarray}
 X(F)&=&\oint \frac{dz}{2\pi i} F(z)X^9(z).
\label{eq:ointFX}
\end{eqnarray}
Therefore, in the NS action around $\Psi_J$ (\ref{eq:action'}),
a field redefinition:
\begin{eqnarray}
 \Phi&=&e^{\frac{i}{2\sqrt{\alpha'}}X(F)}\Phi'
=e^{\frac{i}{2\sqrt{\alpha'}}X_L(F)I}*\Phi'
*e^{-\frac{i}{2\sqrt{\alpha'}}X_L(F)I},
\label{eq:Phi_eXPhi'}
\end{eqnarray}
where
\begin{eqnarray}
 X_L(F)&=&\int_{C_{\rm L}} \frac{dz}{2\pi i} F(z)X^9(z),
\end{eqnarray}
gives the undeformed NS action with respect to a string field $\Phi'$.
In this sense, the solution $\Psi_J$ (\ref{eq:PsiJdef}) induces
 a field redefinition (\ref{eq:Phi_eXPhi'}).

Let us consider the effect of (\ref{eq:Phi_eXPhi'}) for a gauge
invariant overlap.
In order to do that, we take
\begin{eqnarray}
 V_{\rm m}(i,-i)=e^{\frac{i}{2}k_9X^9(i)}e^{-\frac{i}{2}k_9X^9(-i)}
\label{eq:Vm=e^kX^9}
\end{eqnarray}
in (\ref{eq:ginvdef}), where the on-shell condition:
$(k_9)^2=2/\alpha'$ is satisfied,
and it corresponds to a closed tachyon vertex
for a Dirichlet direction.
Furthermore, noting $F(-1/z)=z^2F(z)$,
we expand $F(z)$ as
\begin{eqnarray}
 F(z)=\sum_{m=1}^{\infty}f_m(z^{-m}+(-1)^{m+1}z^m)z^{-1}.
\end{eqnarray}
Then, we have
\begin{eqnarray}
&&\langle I|{\cal V}(i)X(F)=
-4\alpha'k_9\sum_{n=1}^{\infty}\frac{(-1)^nf_{2n-1}}{2n-1}
\langle I|{\cal V}(i)
=2\pi\alpha' k_9\int_{C_{\rm L}}\frac{dz}{2\pi i}F(z)
\langle I|{\cal V}(i),~~~~
\label{eq:IVX=}
\end{eqnarray}
because the $X^9$ sector of $\langle I|{\cal V}(i)$ is 
proportional to \cite{Katsumata:2004cc, Kawano:2008ry}
\begin{eqnarray}
\langle 0|\exp\left(-\sum_{n=1}^{\infty}(-1)^n\frac{1}{2n}(\alpha_n^9)^2
-\sum_{n=1}^{\infty}\frac{2i\sqrt{2\alpha'}(-1)^n}{2n-1}k_9\alpha^9_{2n-1}
\right).
\end{eqnarray}
Using (\ref{eq:IVX=}), the gauge invariant overlap for $\Phi$
(\ref{eq:Phi_eXPhi'}) is evaluated as
\begin{eqnarray}
\langle\!\langle \Phi\rangle\!\rangle_{\cal V}
&=&\frac{1}{2}{\rm Tr}(\sigma_3\langle I|{\cal V}(i)\,
e^{\frac{i}{2\sqrt{\alpha'}}X(F)}|\Phi'\rangle)
\nonumber\\
&=&
\exp\left(i\pi\sqrt{\alpha'}k_9\int_{C_{\rm L}}\frac{dz}{2\pi i}F(z)
\right)
\langle\!\langle
\Phi'\rangle\!\rangle_{\cal V}.
\label{eq:ginv_phase}
\end{eqnarray}
Namely, the solution $\Psi_J$ (\ref{eq:PsiJdef}) induces 
a phase factor in the above gauge invariant
overlap.
In the case of bosonic SFT, such an effect was derived in
\cite{Katsumata:2004cc}.


\end{document}